\documentclass[useAMS]{mn2e}
\usepackage{lscape}
\usepackage{rotating}
\usepackage{amssymb}
\usepackage{float}
\usepackage{hyperref}
\usepackage{newtxtext,newtxmath}

\def\msun{\hbox{M$_\odot$}}

\def\t4{\hbox{t$_{\rm 4}$}}

\def\cm3{\hbox{cm$^{-3}$}}

\input psfig.sty
\input epsf.sty
\usepackage{graphicx}
\usepackage{epsfig}
\title[MPs in Integrated Light]
{Multiple Populations in Integrated Light Spectroscopy of Intermediate Age Clusters}
\author[Bastian et al.] {Nate Bastian$^1$, Christopher Usher$^1$, Sebastian Kamann$^1$, Carmela Lardo$^2$, S{\o}ren S. Larsen$^3$,  \newauthor Ivan Cabrera-Ziri$^4$\thanks{Hubble fellow}, William Chantereau$^1$, Silvia Martocchia$^{1,5}$, Maurizio Salaris$^1$,  \newauthor Ricardo P. Schiavon$^1$, Randa Asa'd$^6$, and Michael Hilker$^5$\\
$^{1}$Astrophysics Research Institute, Liverpool John Moores University, 146 Brownlow Hill, Liverpool L3 5RF, UK\\
$^{2}$Laboratoire d'astrophysique, Ecole Polytechnique F\'ed\'erale de Lausanne (EPFL), Observatoire de Sauverny, CH-1290 Versoix, Switzerland\\
$^{3}$Department of Astrophysics/IMAPP, Radboud University, P.O. Box 9010, 6500 GL Nijmegen, The Netherlands\\
$^{4}$ Harvard-Smithsonian Center for Astrophysics, 60 Garden Street, Cambridge, MA 02138, USA\\
$^{5}$European Southern Observatory, Karl-Schwarzschild-Stra\ss e 2, D-85748 Garching bei M\"unchen, Germany\\
$^{6}$Physics Department, American University of Sharjah, P.O. Box 26666, Sharjah, UAE\\
}
\date{Accepted. Received; in original form}
\pagerange{\pageref{firstpage}--\pageref{lastpage}}
\pubyear{2019}
\begin{document}
\maketitle
\label{firstpage}
\begin{abstract}
The presence of star-to-star light-element abundance variations (a.k.a. multiple populations, MPs) appears to be ubiquitous within old and massive clusters in the Milky Way and all studied nearby galaxies.  Most previous studies have focussed on resolved images or spectroscopy of individual stars, although there has been significant effort in the past few years to look for multiple population signatures in integrated light spectroscopy.  If proven feasible, integrated light studies offer a potential way to vastly open parameter space, as clusters out to tens of Mpc can be studied.  We use the NaD lines in the integrated spectra of two clusters with similar ages ($2-3$~Gyr) but very different masses, NGC~1978 ($\sim3\times10^5$~\msun) in the LMC and G114 ($1.7\times10^7$~\msun) in NGC~1316.  For NGC~1978, our findings agree with resolved studies of individual stars which did not find evidence for Na spreads.  However, for G114, we find clear evidence for the presence of multiple populations.  The fact that the same anomalous abundance patterns are found in both the intermediate age and ancient GCs lends further support to the notion that young massive clusters are effectively the same as the ancient globular clusters, only separated in age. 
\end{abstract}

\section{Introduction}
\label{sec:intro}

Resolved studies of ancient globular clusters (GCs) in the Milky Way (MW) and nearby galaxies, both photometrically and spectroscopically, have shown that all display star-to-star light-element abundance variations within them, known as multiple populations (MPs).  The origin of these variations is still unknown, in part due to the restricted parameter space where such studies are possible.  In the Galaxy, the GC population spans a wide range of metallicities but is limited in the age range that can be probed.  The Large and Small Magellanic Clouds offer cluster populations with a wider range of ages, but a more limited range of masses (especially at younger ages).  Resolved studies with the necessary precision to probe MPs are not currently possible outside the Local Group, and are even severely limited in M31 and M33.  Opening up studies of multiple populations to an increased volume in the local universe would allow access to a much larger portion of parameter space.  This is particularly important as recent studies have suggested that cluster mass (e.g., Carretta et al.~2010; Schiavon et al. 2013; Milone et al. 2017), age (e.g., Martocchia et al. 2018a; 2019) and metallicity (Pancino et al. 2017) all play a role in the manifestation of multiple populations within clusters.

One way to search for MPs in integrated light is to estimate the average abundance of various elements within clusters, and compare the derived mean abundance pattern to expectations based on resolved abundance work in MW GCs (e.g., Colucci et al. 2012; 2014; Sakari et al.~2013; 2016; Larsen et al. 2012; 2017; 2018).  While each GC that has been studied in detail differs in the exact nature of its MPs, there are broad trends that can be used to infer their presence (cf. Gratton et al.~2012).  In particular, the expectation is that the mean abundance of N, Na, and Al should be larger than scaled-solar abundances, while C, O and potentially Mg should be somewhat depleted relative to the field stars of similar metallicity.  Most integrated light studies to date that have searched for MPs have used relatively weak atomic lines at high or medium spectral resolution, which limits such studies to nearby galaxies (a few Mpc) for ancient GCs (e.g., Larsen et al.~2017) and 10s of Mpc for brighter young clusters whose integrated light are dominated by red supergiants (Cabrera-Ziri et al.~2016; Lardo et al.~2017).

Here, we use an alternative method, which has been successfully applied to integrated light studies of galaxies (e.g., Jeong et al.~2013) and ancient globular clusters in M31 (Schiavon et al. 2013), and apply it to two intermediate age ($2-3$~Gyr) massive clusters.  We use the NaD lines to infer whether Na enhancement is present in the clusters.   In intermediate and old stellar populations, NaD is one of the strongest spectral features in the optical region of the spectrum.  This technique has been applied to early type galaxies (e.g., Jeong et al.~2013) which showed that in the absence of interstellar medium (ISM) absorption, the NaD line strength depends primarily on the Na abundance and only weakly on the form of the stellar initial mass function.  For this preliminary study, we use medium and high spectral resolution observations so that we can resolve any ISM absorption in our cluster spectrum, which is expected to have widths of $1-2$~km/s (although this may be more complex in a post-merger system with a highly filamentary dust/gas distribution).  We confirm that in the absence of significant ISM absorption, low-resolution spectra can also be used.

For our study we select two massive clusters, NGC~1978 in the Large Magellanic Clouds and NGG~1316:G114 in the galactic merger remnant NGC~1316, at a distance of 22.9~Mpc (as adopted in Bastian et al.~2006).  NGC~1978 has been studied with high/medium resolution resolved star spectroscopy by Mucciarelli et al.~(2008), who found little or no abundance variations in Na.  Martocchia et al.~(2018a,b) have studied the cluster with HST imaging and found evidence for multiple populations in the form of a relatively small spread in N.  These authors found an age of $2.2$~Gyr, [Fe/H]$=-0.5$, and based on the sub-giant branch could rule out a significant age spread ($<20$~Myr) within the cluster.   G114 in NGC~1316 has previously been the focus of multiple studies and has an age of $2.9\pm0.8$~Gyr, solar metallicity (Goudfrooij et al.~2001) and a dynamical mass of $1.6\times10^7$~\msun\ (Bastian et al.~2006).  Both clusters have low extinction values, with $A_V(G114)$=0 (Bastian et al.~2006; \S~\ref{sec:obs}) and $A_V$(NGC~1978)$=0.22$ (Martocchia et al.~2018). The main cluster properties are given in Table~\ref{tab:properties}.

This paper is organised as follows: in \S~\ref{sec:obs} we introduce the observations and models used, while in \S~\ref{sec:results} we present our main results and discuss their implications.  We present our conclusions in \S~\ref{sec:conclusions}.

\section{Observations and Models}
\label{sec:obs}

\subsection{Observations}

We use the same VLT Ultraviolet and Visual Echelle Spectrograph (UVES) spectrum of NGC~1316:G114 that was used to measure the velocity dispersion of the cluster in Bastian et al.~(2006 - Program ID: 073.D-0305(B)).  We refer the interested reader to that paper for details of the data reduction and spectral details.  Briefly, the wavelength range covered was $4200 - 6200$\AA\ at a resolution of $5$~km/s at $5200$\AA\ ($R\sim60,000$) with a total on-target exposure time of 8.67 hours for G114.  The S/N of the spectra in the NaD region (per \AA) is $\sim150$.  The large radial velocity of NGC~1316:G114 means that the cluster spectral features are redshifted out of the range where Galactic ISM absorption is a potential problem. 

Additionally, a large mosaic of NGC 1316 was observed with VLT/MUSE in program 094.B-0298 (PI: Walcher). The cluster G114 in visible in the central two pointings of the mosaic, which we reduced and combined using the standard MUSE pipeline (Weilbacher et al. 2012; 2014). The final data cube consists of 6 individual exposures, with an integration time of 150s each. G114 is visible in all exposures, hence the effective exposure time for the cluster is 900s. The average seeing in the final data cube is 0.8".  The spectrum of G114 was extracted by summing the fluxes of all spaxels within a distance of $0.8\arcsec$ from the visually determined cluster location. The contribution of the host galaxy was accounted for by averaging the fluxes of all spaxels in an annulus between $1.6\arcsec$ and $2.8\arcsec$ distance from said location and subtracting the resulting spectrum from the spectrum extracted for G114. We verified that the final result did not depend sensitively on the extraction radii that were used.  The extracted MUSE spectrum has S/N (per \AA) of $30$ in the NaD region of G114. 

For NGC~1978 we use the integrated light spectrum from the WAGGS project (WiFeS Atlas of Galactic Globular cluster Spectra).  The observations and reductions are described in detail in Usher et al.~(2017).  In short, the spectrum covers the wavelength range from $3300$\AA\ to $9050$\AA, with a resolution of $R=6800$, and a S/N in the NaD region of 80 (per \AA).  The cluster radial velocity is high enough to separate the absorption from the Milky Way from the cluster, as can be seen in Fig.~\ref{fig:nad}.

Finally, in order to verify the low extinction towards G114, we downloaded images of NGC~1316 from the Hubble Legacy Archive, which use the Advanced Camera For Surveys, Wide Field Camera in F435W, F555W and F814W filters (GO-9409; PI P. Goudfrooij - see Goudfrooij et al.~2012 for a discussion of the data).  Aperture photometry was carried out on G114 with the resulting colours of {\em F435W-F814W=1.90} and {\em F555W-F814W=1.03}.  Comparing the measured colours (for {\em F555W-F814W}) to those of Goudfrooij et al. (2012) shows excellent agreement.  Comparing these colours to expectations of a $3$~Gyr, solar metallicity SSP from Bruzual \& Charlot (2003; 2016 edition), which have 2.01 and 1.2 for {\em F435W-F814W} and {\em F555W-F814W}, respectively, shows good agreement.  Both model colours are $0.1$~mag bluer for lower (half-solar) metallicity.  The fact that the observed colours are similar to, although slightly bluer than, the models confirms the low extinction towards G114.

\begin{table} 
  \begin{tabular}
    {llllll} Galaxy & Cluster & Age & Mass & $\sigma$ & v$_{\rm r}$\\
    & & (Gyr) & (\msun) & (km/s) & (km/s) \\
    \hline
 LMC & NGC~1978 & 2.2 & $3\times10^5$ & 3.1 & 293\\
 NGC~1316 & G114 & 3 & $1.6\times10^7$ & $42$ & 1292\\
 \hline
  \end{tabular}
\caption{Properties of the clusters studied in the present work.  NGC~1978 has [Fe/H]$=-0.5$ (Martocchia et al.~2018), while G114 has solar metallicity (Goudfrooij et al. 2001).  References for the values are given in the text.}
\label{tab:properties}
\end{table}

\subsection{Stellar Population Models}

The models were developed in the same way as the integrated light models of early type galaxies presented in Chantereau et al. (2018), which were based on the stellar models used in Martocchia et al.~(2017).  We refer the interested reader to these papers for more details.  In summary, we use the {\sc MIST} isochrones (Dotter et al. 2016, Choi et al. 2016), at a given age and metallicity, to sample the distribution of luminosity and effective temperature of stars at all evolutionary stages.  Stellar model atmospheres were computed for each selected point in the HR-diagram using {\sc ATLAS12} (Kurucz 1970; 2005) and synthesised stellar spectra were created with {\sc SYNTHE}  (Kurucz \& Furenlid~1979; Kurucz \& Avrett~1981).  For initial abundances of the stars we adopted three compositions, with the first being scaled solar.  For the other two we adopted abundance patterns representative of multiple populations seen in Galactic globular clusters (e.g., Sbordone et al.~2011), although note that we did not use $\alpha$-enhanced abundances.  The first of these, which we will identify as 'intermediate' uses [N/Fe] = [Na/Fe] = [Al/Fe] = 0.5, [C/Fe] = [O/Fe] = -0.09, and [Mg/Fe] = -0.07.  The second of these, termed 'extreme' adopts  [N/Fe] = [Na/Fe] = [Al/Fe] = 1.0, [C/Fe] = [O/Fe] = -0.70, [Mg/Fe] = -0.44.  These values were chosen to  keep the C+N+O, Ne+Na and Mg+Al sums constant.  Both the isochrones and the spectral synthesis adopted the solar scaled abundance of Asplund et al.~(2009, Y = 0.27, Z = 0.0142) with [Fe/H]$=0.0$ for G114 and $-0.5$ for NGC~1978.

We note that the CNONaMgAl variations were applied only to the spectra, hence the mass, L and $T_{\rm eff}$ distributions of scaled solar isochrones remain unchanged.  This is a justified assumption as long as the CNO sum remains unchanged (e.g., Cassisi et al. 2013), which is true in our models.  We adopted a Kroupa~(2001) stellar initial mass function with a lower mass limit of $0.08$\msun.  However, the lowest mass model spectra that was produced was for $0.2$~\msun.  We checked the impact on the NaD lines in our models by the adopted initial mass function (comparing a Kroupa~2001 to a Salpeter~1955 distribution) and the differences were less than a few percent, consistent with what has been found by other authors (e.g., Conroy \& van Dokkum~2012; Jeong et al.~2013).

The models were initially computed for R=200,000 and were then convolved with the velocity dispersion or resolution measured for each of the clusters $\sigma_{\rm G114}=42$~km/s (Bastian et al. 2006) and R=6800 for NGC~1978 (Usher et al.~2017).  We computed specific models for each of the clusters, adopting the parameters for age and metallicity as discussed in \S~\ref{sec:intro}.  The impact of adopting models with different ages and metallicities is explored in Appendix~A.

One of the drawbacks of using integrated light spectroscopy is that it suffers from a degeneracy between the severity of the abundance variations (i.e., how enhanced or depleted a given element is) and the fraction of stars that display the anomalies (e.g., Schiavon et al. 2013).  In the Galaxy, we see that the fraction of stars that display the chemical anomalies increases with increasing present day cluster mass (Milone et al.~2017) and also the extent of the variations (e.g., the maximum spread seen in a given element within the cluster) also varies with increasing mass (e.g., Bastian \& Lardo~2018).  Using integrated spectroscopy, we only get a (light-weighted) mean abundance for the full population for each element studied.

\section{Results}
\label{sec:results}

\subsection{High and medium resolution spectroscopy}
\begin{figure}
\centering
\includegraphics[width=8cm]{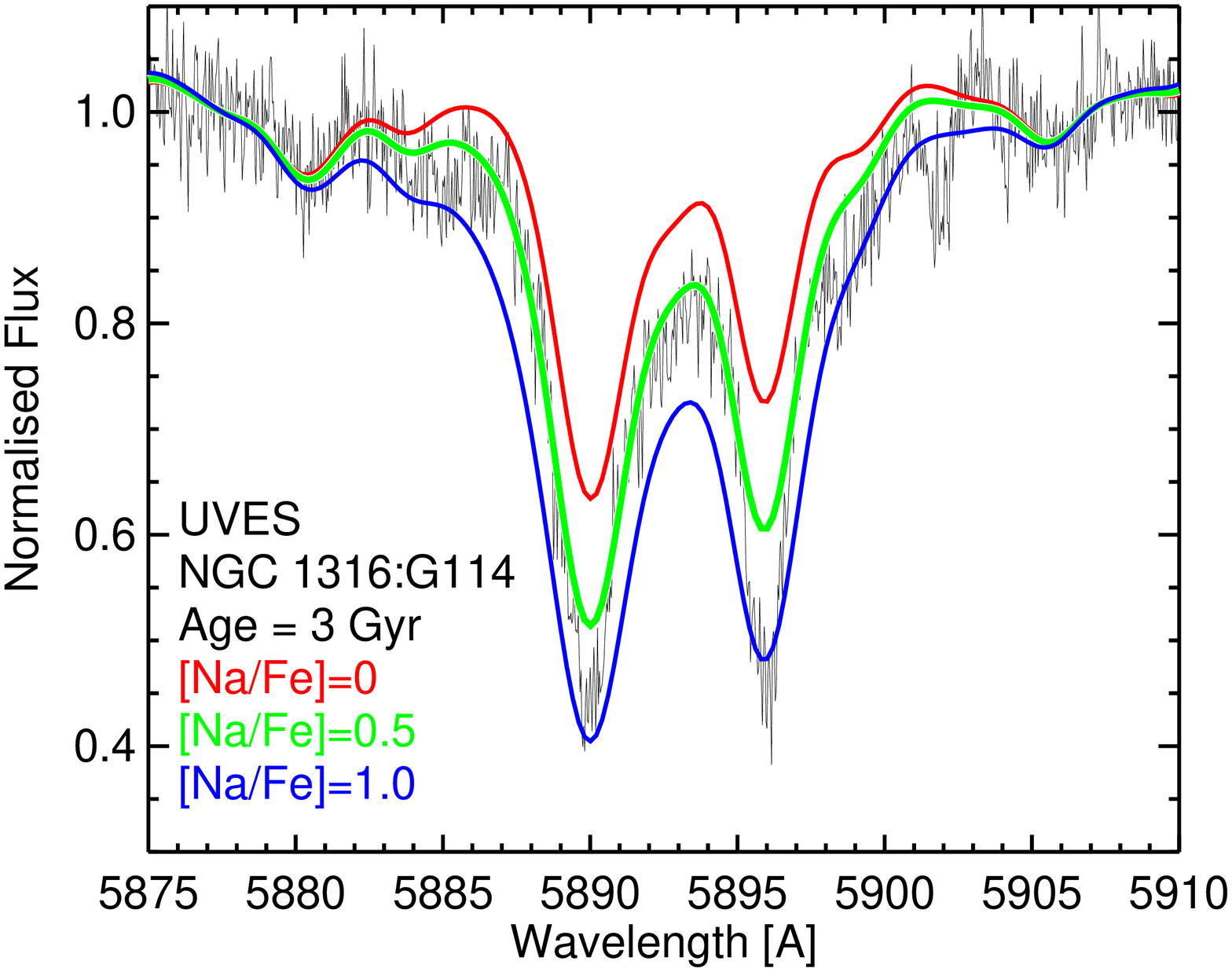}
\includegraphics[width=8cm]{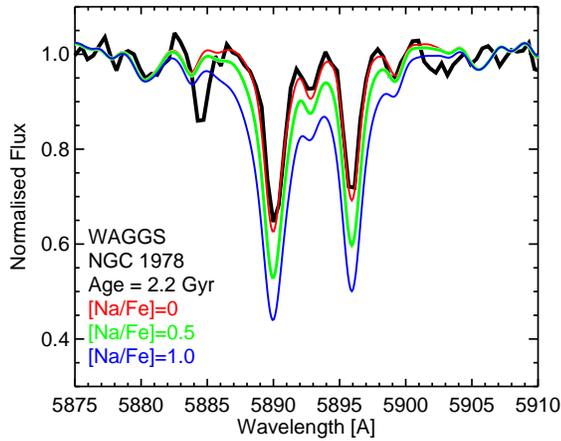}
\caption{{\bf Top:} The NaD line of the $\sim3$~Gyr cluster, NGC~1316:G114 is shown as a solid black line, along with integrated stellar population models with different levels of multiple populations, from solar abundance ratios (red) to intermediate (green) and extreme (blue). {\bf Bottom:} The same as the top but now for the $\sim2.2$~Gyr cluster, NGC~1978 in the LMC.}
\label{fig:nad}
\end{figure}

In Fig.~\ref{fig:nad} we show the spectral region of NaD for NGC~1316:G114 (top) and NGC~1978 (bottom), along with three models for solar scaled (red), intermediate (green), and extreme (blue) abundance patterns and we list the corresponding Na enhancement in the legend.  We see that the solar scaled abundance models predicted that NaD would be much weaker than observed for G114.  The observations are best reproduced (in terms of line profile) with the intermediate abundance models. Interpolating from the models shown, we find that G114 has an [Na/Fe] value of $0.6\pm0.1$, based on fitting the line profile.  This is our primary evidence that Na is enhanced within this cluster, hence that it contains multiple populations.  We note that the cores of the lines in G114 are observed to be slightly stronger than predicted by the best fitting model, which may indicate the presence of NLTE effects that are not taken into account in the modelling (Mashonkina et al.~2000).

\begin{figure}
\centering
\includegraphics[width=8cm]{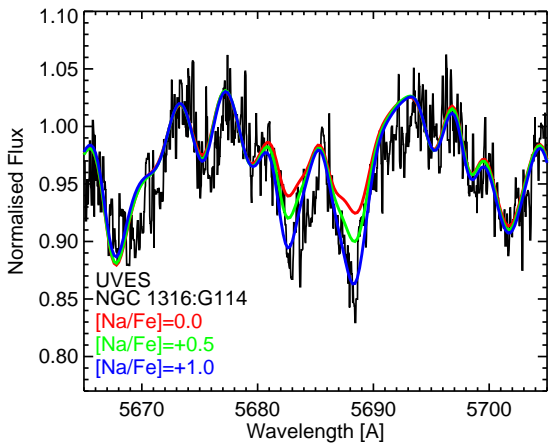}

\caption{Spectral region of NGC~1316:G114 including the Na doublet at 5683\AA\ and 5688\AA.  The observed spectra have been box-car smoothed by 5 pixels for clarity.  These lines are not affected by ISM absorption.}
\label{fig:na}
\end{figure}

We look for validation of this result using the weaker Na lines at $5683$\AA\ and $5888$\AA, which are not affected by ISM absorption and where NLTE effects are expected to be weaker.  The results for G114 are shown in Fig.~\ref{fig:na}.  While the signal-to-noise in these lines is much lower than in the NaD lines, we see again that the observed lines are stronger than would be expected based on solar-scaled abundances, hence that Na is enhanced.  In this case the best fitting model is closer to the extreme abundance pattern, although due to the lower S/N here than in the NaD region we consider the two measurements to be consistent.

In contrast to G114, the spectrum of NGC~1978 is well reproduced by the solar-scaled abundance of Na (at LMC metallicity).  This is in agreement with high resolution spectroscopy of individual stars that also found little or no Na spread within this cluster (Mucciarelli et al.~2008).  While this cluster does host a spread in N (Martocchia et al.~2018a,b) it does not in Na, at least not above the measurement uncertainties.  The fact that we do not find enhanced Na in this cluster lends further support to our analysis method, and hence to the conclusion that there is a significant Na spread within G114.

\subsection{Low resolution spectroscopy of G114}

\begin{figure}
\centering
\includegraphics[width=7.5cm]{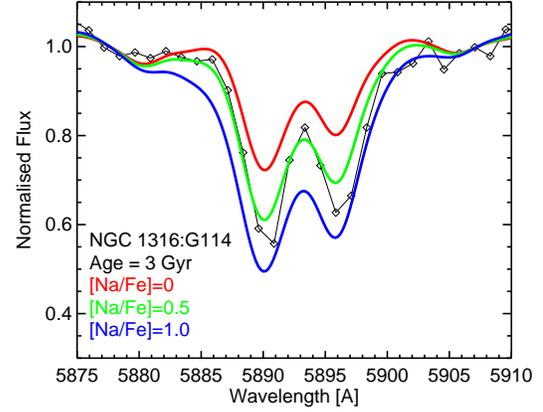}
\caption{The NaD spectral features for NGC~1316:G114 observed with MUSE.  The models are for an age of $3$~Gyr and [Fe/H]$=0.0$ for different levels of MPs, convolved to match the spectral resolution of MUSE. }
\label{fig:g114_muse}
\end{figure}

We have also analysed the NaD line in the extracted MUSE spectrum of NGC~1316: G114 and the results are shown in Fig.~\ref{fig:g114_muse}, along with the three MP models (smoothed to the MUSE resolution).  As was found with the UVES spectrum, the observations suggest the presence of MPs with the intermediate model (or slightly more enhanced) providing the best fit.  Hence, in the absence of strong ISM absorption, low resolution spectra are also able to be used to determine whether Na spreads are present.   

In addition to Na, there are a number of elements that are seen to vary within globular clusters.  One such element is Al which is seen to vary, to first order, in an equivalent way and amount as Na (e.g., Carretta et al. 2009b).  There are a few of Al lines in the red part of the optical spectrum that, in principle, can be used to either confirm or refute the Na spreads found in the present work.  The line(s) with the strongest difference between the solar-scaled (i.e., primordial) abundance and the intermediate or extreme abundance patters are two lines close together at 6696.01\AA \& 6698.67\AA, that appear as a single line in low resolution spectra.  

\begin{figure}
\centering
\includegraphics[width=7.5cm]{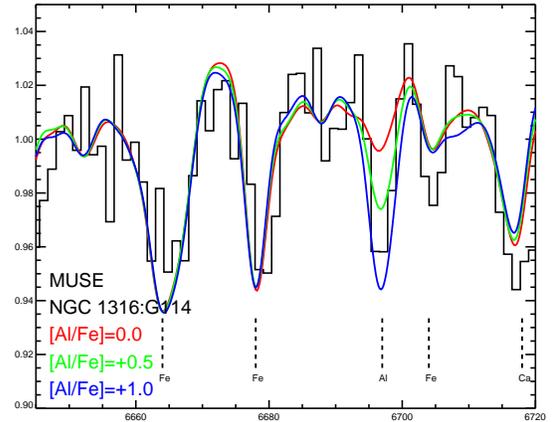}
\caption{The portion of the MUSE spectra of NGC~1316:G114 of the Al lines near 6697\AA.  Additionally, we show the primordial, intermediate and extreme SSP models convolved to the MUSE resolution.  There is tentative evidence for Al enhancement, further strengthening the argument the MPs are present in this cluster.}
\label{fig:al}
\end{figure}

We show this portion of the MUSE spectra for NGC~1316:G114 in Fig.~\ref{fig:al}.  We see some evidence for [Al/Fe] to be enhanced, a further indication that MPs are present in this cluster.  However, we note that, even for the strongest of the Al lines, the lines are relatively weak and model uncertainties within this region (and at this difference level) make this detection tentative.  Higher S/N, and possibly higher resolution, spectra will be required to confirm this.

Unfortunately, the S/N of the WAGGS NGC~1978 spectrum is too low to carry out an equivalent test on that cluster.

\section{Discussion and Conclusions}
\label{sec:conclusions}

We have presented a technique to identify and quantify the presence of multiple populations in the integrated spectra of massive stellar clusters, that is, at least in principle, applicable at any age.  The technique makes use of the strong NaD lines, along with dedicated stellar population modelling, to look for enhanced [Na/Fe].  While the NaD lines can be affected by ISM absorption, this effect can be mitigated by using high resolution spectroscopy (as ISM absorption is typically very narrow) or by studying clusters with little or no foreground extinction.   We have applied the technique to two clusters with similar ages, NGC~1978 ($2.2$~Gyr) and NGC~1316:G114 ($\sim3$~Gyr), but with very different masses, $3\times10^5$~\msun\ and $1.7\times10^7$~\msun, respectively.

NGC~1978, in the LMC, has been studied previously, using high-resolution spectroscopy of individual stars, where little or no spread in [Na/Fe] was found (Mucciarelli et al.~2008).  Using our technique, we come to the same conclusion, namely that there is no evidence for significant Na enhancement within this cluster.  For NGC~1316:G114, the much more massive cluster, we find clear evidence for the presence of multiple populations, as Na appears to be quite enhanced, with [Na/Fe]$=0.6\pm0.1$~dex.

While MPs have been detected in NGC~1978 previously, based on an inferred spread in nitrogen (Martocchia et al. 2018a), the corresponding spread in Na is much smaller (or non-existent).  One potential cause for this apparent contradiction is that mass is known to play a key role in the manifestation of MPs (e.g., Carretta et al.~2010; Schiavon et al.~2013; Milone et al.~2017).  We note that the ancient GCs in the MW, with masses comparable to NGC~1978, typically do have detectable Na spreads ($>0.5$~dex; Carretta et al.~2009), hence age may also be playing an important role (Martocchia et al.~2018a).

We note that the discovery of MPs in intermediate age clusters was, until now, based primarily on N variations, hence the evidence for Na variations reported in the present work suggests that many of the MP abundance variation signatures are present (i.e., multiple elements, not just N).  Future higher S/N spectra as well as near-UV and/or near-IR observations of the two clusters studied here should be able to determine if other elements, such as N and O are varying as expected. 

The [Na/Fe] enhancement inferred for NGC~1316:G114 is larger than that seen in the majority of integrated studies of ancient GCs (e.g., Larsen et al. 2017; 2018).  As resolved studies have shown that abundance variations become more extreme in higher mass clusters (at least in the studies limited to ancient clusters), a reasonable interpretation of this result is that due to the extreme mass of this cluster ($\sim100\times$ that of typical Galactic GCs) its abundance variations are correspondingly stronger.  Hence, our results support the previous suggestions that cluster mass, and potentially age, play important roles in setting the (observed) MP properties of clusters.

The technique used here has the potential to significantly open up parameter space in the study of multiple populations.  By studying the integrated spectra of other high mass clusters formed in galactic mergers and starbursts with ages between $\sim100$~Myr and $5-6$~Gyr we should be able to better understand the dependence of MPs on cluster age and mass.  By looking at clusters within the same galaxy that have similar ages, we will be able to study the evolution of the correlation between cluster mass and MP properties. The technique also offers a way to study MPs in ancient globular clusters when the S/N is not high enough to study the weaker atomic lines. 

One potential caveat to the method presented here is that it assumes that [Na/Fe] (or other elements studied) is enhanced relative to the host galaxy, i.e. that the observed excess of [Na/Fe] is a feature of MPs and not an overall galactic trend.  This caveat can be mitigated, to a certain extent, by looking at other abundances that are also linked to MPs.  Ideally, one would like the full spectrum of elements to be assessed in both the host galaxy and the cluster under study, in order to see the complete picture of abundance variations.

These results provide further evidence that the multiple populations phenomenon is not restricted to the ancient globular clusters, but seems to be a general characteristic of all massive clusters.  This suggests a common formation mechanism for clusters across all cosmic epochs (see Forbes et al.~2018 for a recent review), i.e. that these massive clusters are truly young globular clusters.

\section*{Acknowledgments}

We thank the anonymous referee for helpful suggestions.  NB gratefully acknowledges financial support from the Royal Society (University Research Fellowship).  NB, CU, and SK gratefully acknowledge financial support from the European Research Council (ERC-CoG-646928, Multi-Pop).  CL acknowledges financial support from the Swiss National Science Foundation (Ambizione grant PZ00P2\_168065).  Support for this work was provided by NASA through Hubble Fellowship grant HST-HF2-51387.001-A awarded by the Space Telescope Science Institute, which is operated by the Association of Universities for Research in Astronomy, Inc., for NASA, under contract NAS5-26555. WC acknowledges funding from the Swiss National Science Foundation under grant P400P2\_183846.

\appendix

\section{Uncertainties regarding G114}
\label{sec:g114_uncertainties}

\begin{figure}
\centering
\includegraphics[width=8cm]{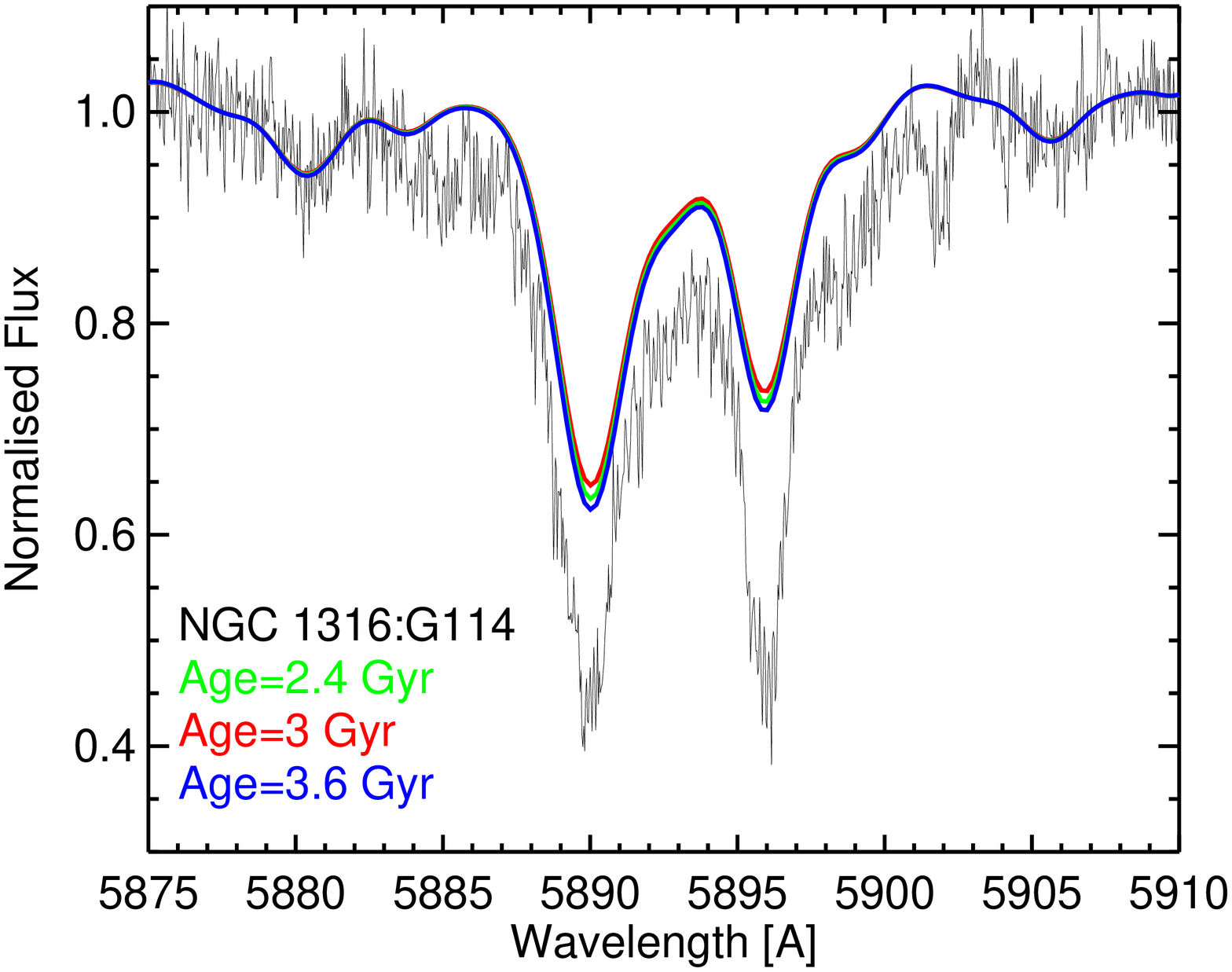}
\includegraphics[width=8cm]{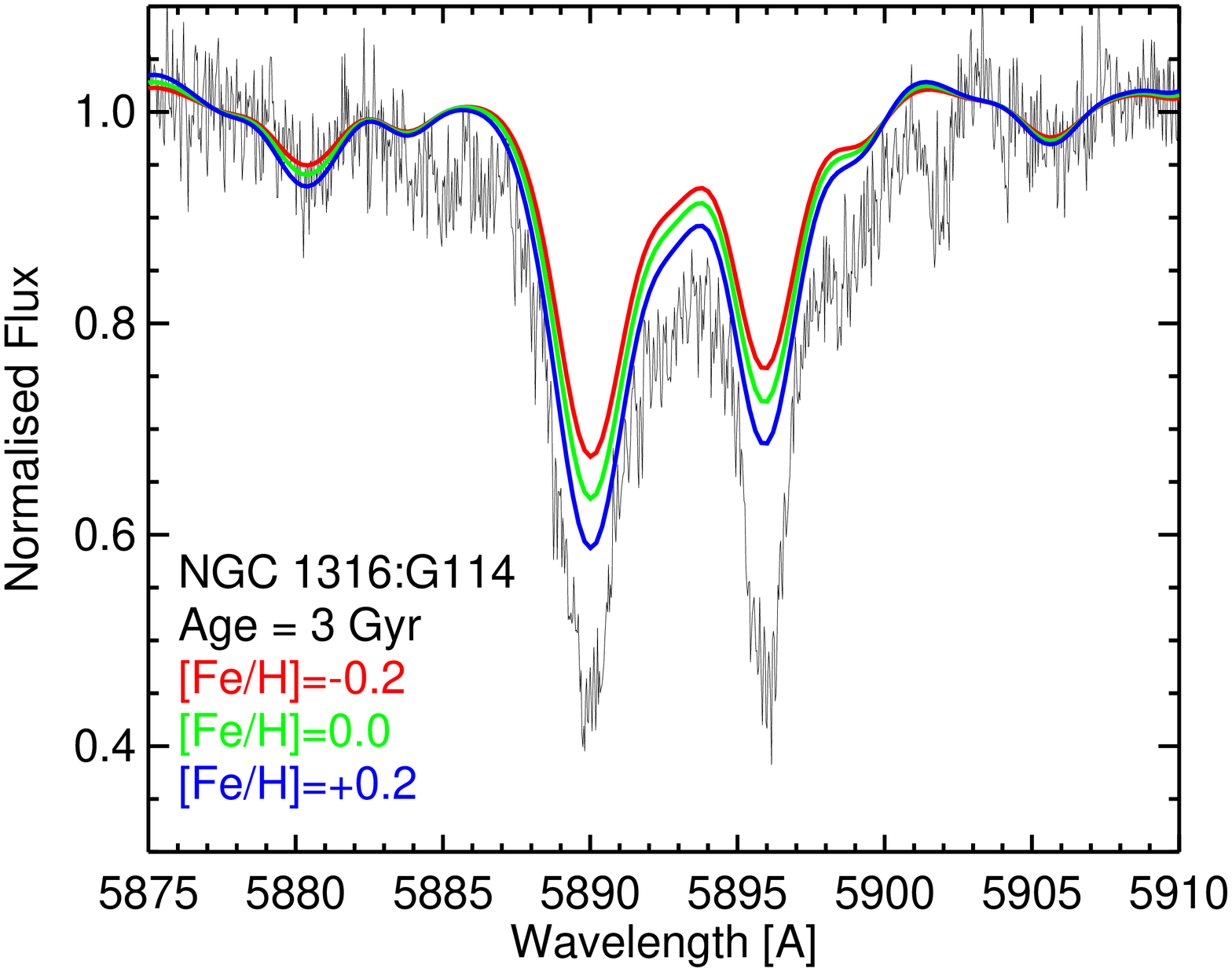}

\caption{The NaD spectral features for G114 observed with UVES for different combinations of model parameters.  {\bf Top:} The effect of uncertainties on the age of G114 (at [Fe/H]=0.0).  {\bf Bottom:} The effect of uncertainties on the metallicity of G114.}
\label{fig:g114_age_z}
\end{figure}

\begin{figure}
\centering
\includegraphics[width=8cm]{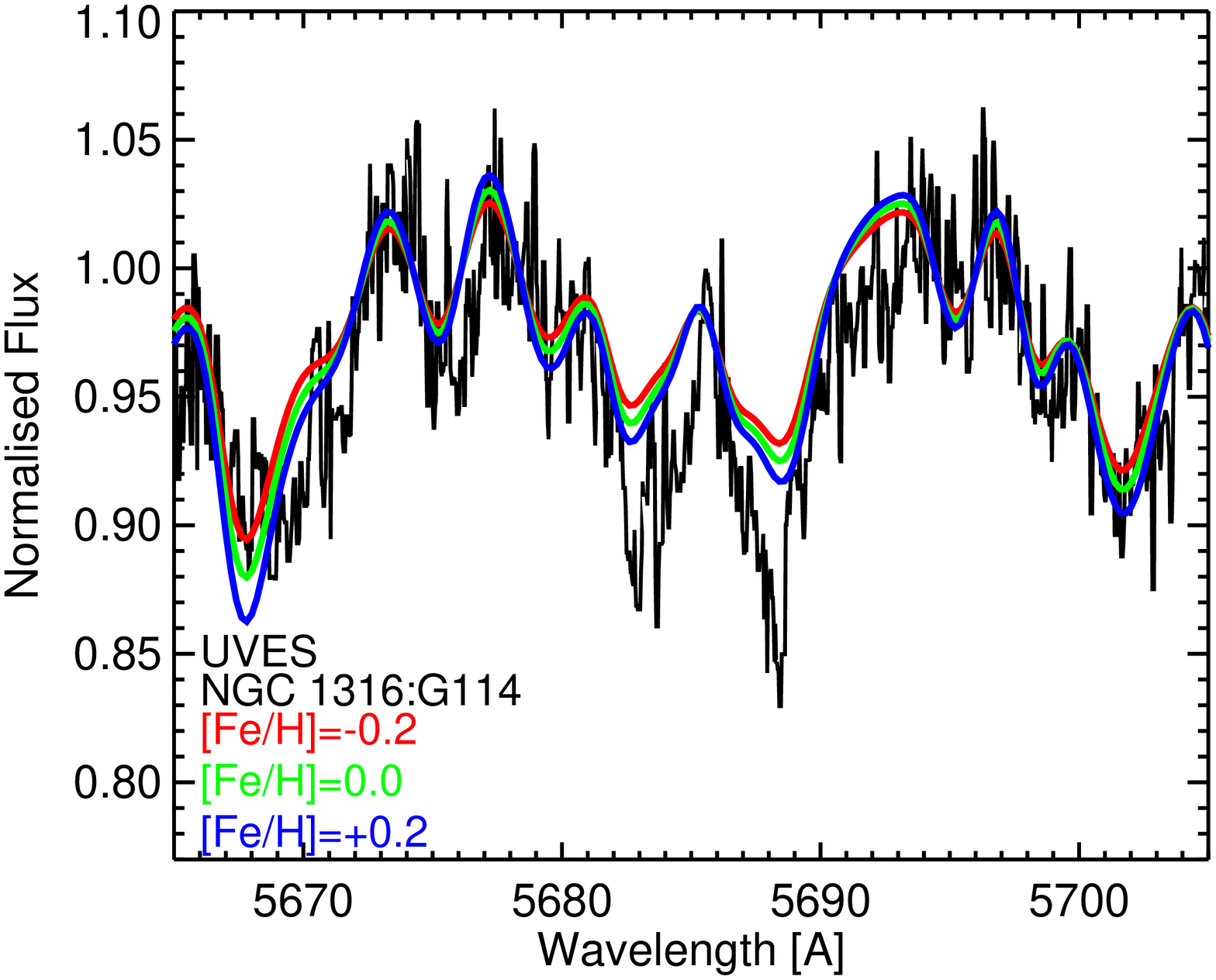}

\caption{The same as the bottom panel of Fig~\ref{fig:g114_age_z} but now for the Na lines at 5683\AA~ and 5688\AA.}
\label{fig:g114_na_z_test}
\end{figure}

We have created a number of synthetic models in an attempt to reproduce the spectrum of G114.  Specifically, we investigate the uncertainties involved if the adopted properties of G114 ($3$~Gyr, solar metallicity) were altered by 20\%.  The results are shown in Fig.~\ref{fig:g114_age_z}.  We find that at an age of $\sim3$~Gyr, the uncertainties in the age and metallicity do not strongly impact our interpretation of whether MPs are present or their strength.  In the bottom panel, we see that [Fe/H]$\gtrsim+0.6$ would be needed to explain the NaD line strength (assuming [Na/Fe] is not changing).  A similar conclusion is reached using the Na lines at 5683\AA~ and 5688\AA, which are shown in Fig.~\ref{fig:g114_na_z_test}.

\begin{figure}
\centering
\includegraphics[width=8cm]{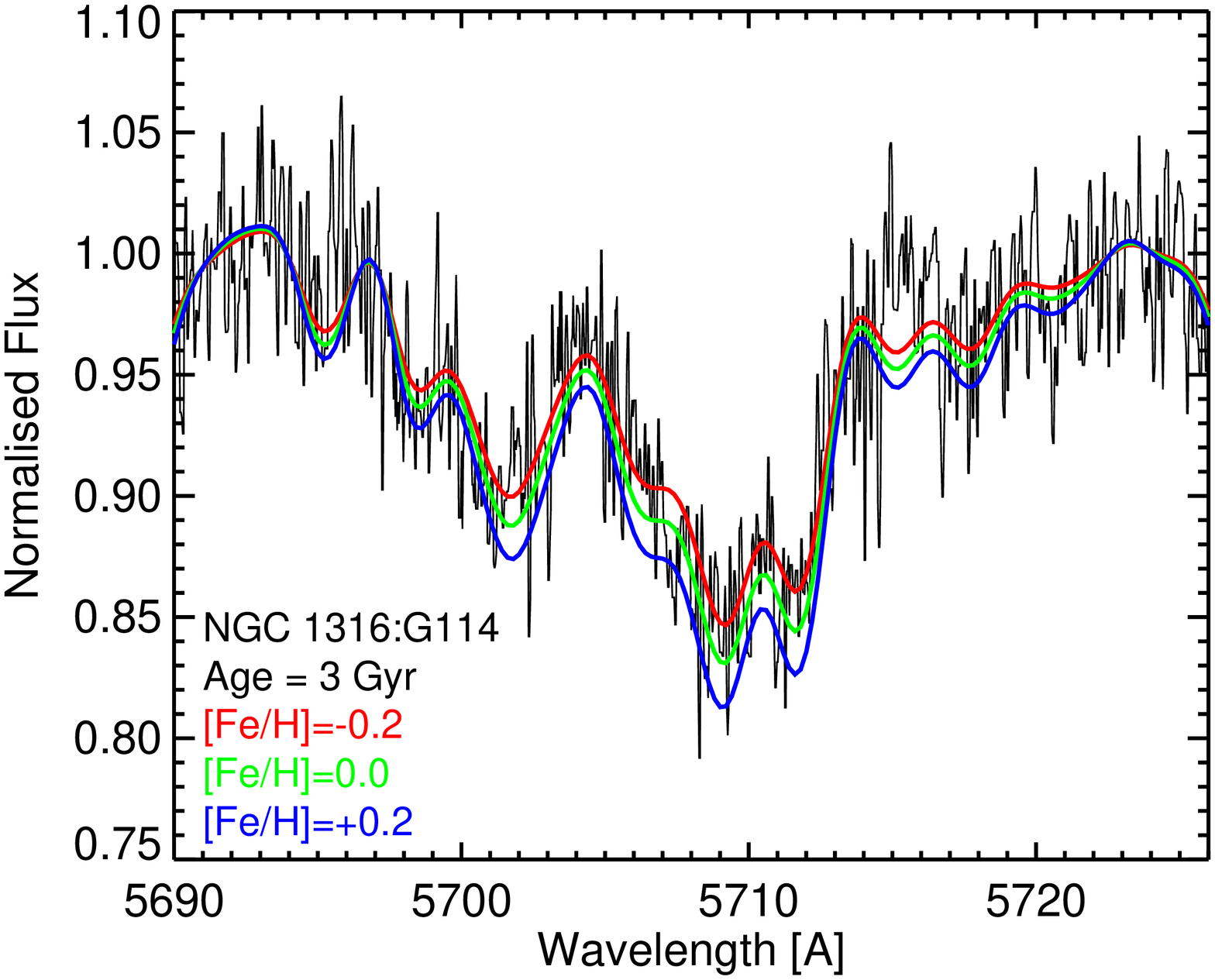}
\includegraphics[width=8cm]{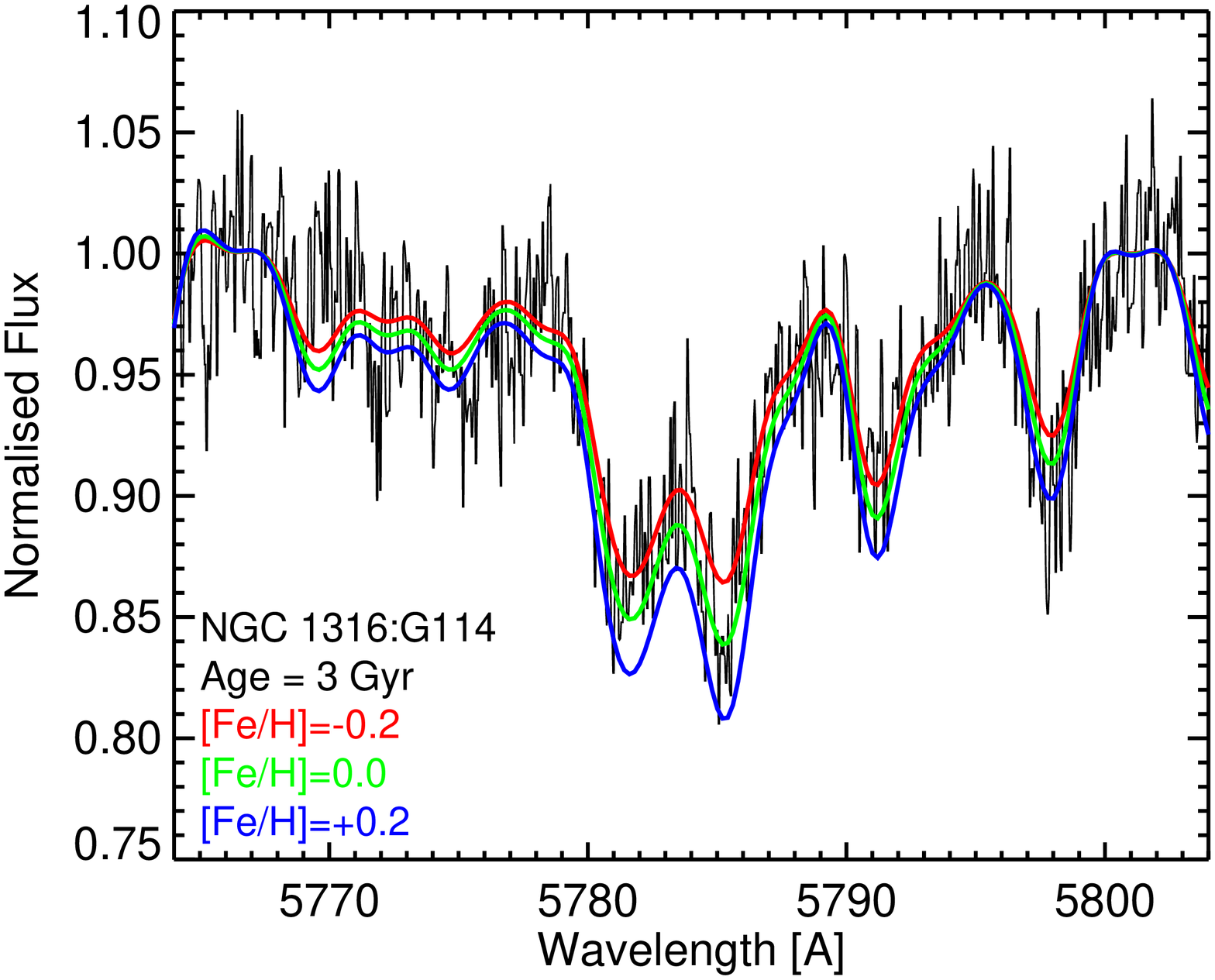}

\caption{Two regions of the UVES spectrum of G114 centred on Fe Lick indices, Fe5709and Fe5782 in the top and bottom panels, respectively.  We also show the same models used in the bottom panel of Fig.~\ref{fig:g114_age_z}, for three different [Fe/H] values (at fixed age).  The data have been boxcar smoothed by three pixels for clarity.  The best fitting model is the [Fe/H]$=0.0$~ model, and the [Fe/H]$=+0.2$~ model is already over-predicting the line strengths. }
\label{fig:g114_feh}
\end{figure}

We have also explored specific spectral regions that are dominated by Fe lines.  This is shown in Fig.~\ref{fig:g114_feh}, where the two panels are centred on the Lick indices Fe5709 and Fe5782.  We  over-plot models with different [Fe/H] values.  We see that for both regions, the [Fe/H]$=0.0$ model best reproduces the observed spectral regions and that the [Fe/H]$=+0.2$ model is over-predicting the line strengths.  We conclude that the adopted [Fe/H] value is accurate to within 20\% and that [Fe/H]$<0.2$ for G114, hence a super-solar metallicity cannot explain the observed strength of the NaD lines.

\bsp
\label{lastpage}
\end{document}